\documentclass[twocolumn,showpacs,preprintnumbers,amsmath,amssymb]{revtex4}
\usepackage{graphicx}
\begin{document}

%
%
\title{Multipulse Control of Decoherence}  
\author{Chikako Uchiyama}
\affiliation{%
Faculty of Engineering,Yamanashi University,
4-3-11, Takeda, Kofu, Yamanashi 400-8511, JAPAN
}%
\author{Masaki Aihara}
\affiliation{%
Graduate School of Materials Science,
Nara Institute of Science and Technology,
8916-5, Takayama-cho, Ikoma, Nara 630-0101 JAPAN
}%
\date{\today} 
%
\def\ch{{\cal H}}
\def\hbar{\mathchar'26\mkern -9muh}
\def\muv{{\vec \mu}}
\def\ev{{\vec H}}
\def\ih{\frac{i}{\hbar}}
\def\taus{{\tilde \tau}_{s}}
\def\tauc{{\tilde \tau}_{c}}

\begin{abstract}
We present a general formulation to suppress pure dephasing by multipulse control.  The formula is free from a specific form of interaction and is expressed in terms of the correlation function of arbitrary system-reservoir interaction.  We first apply the formula to a phenomenological two-level model where the correlation function of the interaction decays exponentially. In this case, we analytically show that the pure dephasing time is effectively lengthened by the multipulse control.  Secondly, we apply the formula to the spin-boson model where a spin nonlinearly interacts with a boson reservoir.  We find the multipulse control works well when the pulse interval is sufficiently shorter than the correlation time of system-reservoir interaction. Moreover, in this case, the pure dephasing can also be suppressed by adjusting the pulse interval to the period of dynamical motion of reservoir.
\end{abstract}

\pacs{03.65.Yz,03.67.Hk,05.30-d}%
\maketitle

\section{Introduction}
\label{sec:1}
The reliability of quantum information rests on the stability of quantum 
superposed state.  Since the quantum superposition is vulnerable to the 
effect of the environment,  the degradation
of the superposed state would affect the accuracy of the quantum information, especially in quantum computation. 

In order to reduce errors caused by the environmental effects, many methods have been proposed. These are roughly divided into two categories whether the methods requires ancillary bits or not.  Since the first proposal by Shor\cite{shor}, considerable works on the methods with additional qubits have been done to overcome the error by distributing the information of a qubit over several qubits \cite{gottesman,laflamme,calderbank,steane,zanardi1,zanardi2} including experimental studies of NMR\cite{cory,viola,fortunato} and trapped ions \cite{kielpinski}.   However, in these methods, it has been a main problem to keep the coherence of all the qubits including ancillary bits, which may be an obstacle for scale up of quantum computer.  

Two distinctive approaches have been taken to attain strategies for control of decoherence without additional bits.  Firstly, closed-loop control (quantum feedback) method has been proposed\cite{vitali1,vitali2,vitali3,vitali4}.  Since they require the measurement on the system to control the error, the quantum detection efficiency gives a limitation to execution of the control.    
The other method to suppress the error by decoherence without ancillary bit has been called the open-loop control\cite{llyod1,ban,luming}.  In the pioneering works of Viola and Lloyd\cite{llyod1}, and Ban\cite{ban}, they have proposed that a pulse train can be used to eliminate pure dephasing of a spin caused by interaction with a quantum reservoir.  They assume that the spin linearly interacts with boson reservoir.  The pure dephasing of superposed state is successfully suppressed by applying many successive short \(\pi\) pulses. An actual evaluation has been done by using the coupling function for the macroscopic tunneling phenomena as proposed by Caldeira and Leggett\cite{Caldeira}.  Although the pulse methods have been well known in the field of NMR to decouple unnecessary spin-spin interaction\cite{ernst}, their formulation can also be applied to suppress decoherence in various systems other than the nuclear spin system.  The applicability has been generally discussed in \cite{lloyd2}.  Application of a \( 2 \pi \) pulse train to a two-level system, which interacts with a boson reservoir by continuous coupling function, has been studied to discuss the controllability of unwanted spontaneous emission\cite{agarwal1,agarwal2}.  The magnetic state decoherence by collisions in a vapor has been analyzed to be suppressed by ultra fast pulses\cite{search1,search2}.  Moreover, the method is applied to the damped vibrational mode of a chain of trapped ions\cite{vitali}.   

It is however noted that, in the analysis \cite{llyod1,ban}, the interaction between the spin and bosons are assumed to be linear in the boson field amplitude.  We should recall that in many systems the dephasing phenomena cannot be described with the linearly interacting spin-boson model, except for the ohmic case of the Caldeira-Leggett model.  This is because the dephasing rate is given by the zero-frequency component of the power spectrum of interaction amplitude \cite{aihara1,aihara2}. A typical example is a localized-electron phonon system in which the phonon with zero frequency (i.e., translation of a whole crystal) does not interact with the \(\frac{1}{2}\) spin (the two-level electron).    The nonlinear interaction between the spin and phonons, where simultaneous multiple absorption and emission of phonons occurs, should be taken into account for the correct description of the dephasing phenomenon.  In order to discuss the effectiveness and generality of the multipulse control of qubits, a spin-boson model that includes nonlinear interaction is necessary. 

In this paper, we present a general formulation free from a specific form of interaction, and apply this to the two cases (1)a phenomenological two-level model where the correlation function of the interaction decays exponentially and (2)a quadratically interacting localized-electron phonon system. 

This paper is composed as follows: In section \ref{sec:2} , a general formula which 
describes spin dynamics under multipulse control is obtained for an arbitrary interaction with the cumulant expansion method. The expression is written with the correlation function of interaction amplitude. In section \ref{sec:3}, we first apply the formula to a simple case of the exponentially decaying correlation function in order to generally discuss the memory effect of the reservoir causing the spin dephasing.    We next apply the formula to a quadratically interacting localized-electron phonon system, and the dynamical behavior of spin is analyzed.  Discussion and concluding remarks for the usefulness of the multipulse control of spin dephasing are given in section \ref{sec:4}.

\section{A General Formula of the multipulse control}
\label{sec:2} 
We consider the pure dephasing phenomena of a
\(\frac{1}{2}\) spin caused by an interaction with a reservoir.
Here and hence force, we use the term ``spin'' as an arbitrary two-level quantum system with energy difference \(\hbar \omega\).
The Hamiltonian of this system is 
\begin{equation}
\ch_{R}=\ch_{0} + \ch_{SB}\;,
\label{eqn:1}
\end{equation}
where \(\ch_{0} \) and \(\ch_{SB}\) are defined by
\begin{eqnarray}
\ch_{0}&=& \hbar \omega S_{z} + \ch_{B} \;, \label{eqn:2} \\
\ch_{SB}&=& \hbar S_{z}B \;. \label{eqn:3}
\end{eqnarray} 
Here  \(S_{z} \) is the z-component of the \(\frac{1}{2}\) spin, \(\ch_{B} \) is a Hamiltonian of the reservoir and \(B\) is an interaction amplitude operator which consists of reservoir operators and causes the energy change in the energy difference between two states due to reservoir.

It is our purpose to study effectiveness of a multipulse control on this spin system.  Assuming that pulse duration is so short that the spin-reservoir interaction is
ignored during pulse excitation, we write the corresponding Hamiltonian \(\ch_{P} \) during a pulse as
\begin{equation}
\ch_{P}(t) =\ch_{0}+W_{P}(t) \;,
\label{eqn:4}
\end{equation}
where \(W_{P}(t) \) denotes:
\begin{equation}
W_{P} (t)= -\frac{1}{2} \ev  \cdot \muv \; (S_{+} e^{-i \omega t}+S_{-} e^{i \omega t}) \;.
\label{eqn:5}
\end{equation}  
Here we consider the case where an applied square pulse of the magnetic field \(\ev\) is on resonance with the spin whose magnetic moment is defined by \(\muv \).  Using an electric field instead of the magnetic field \(\ev\) and a dielectric moment instead of the magnetic moment \(\muv \), we can apply the following formulation to an electronic state in a quantum dot.

The phase information of the spin is obtained by calculating the off-diagonal element of the reduced density operator, \( \langle 1 | Tr_{R} \; \rho(t) | 0 \rangle \) where \(| 1 \rangle (| 0 \rangle )\) is the upper (lower) state of the spin,  \(\rho(t)\) is the density operator of the total system and \(Tr_{R}\) denotes  the trace operation over the degrees of the freedom of the reservoir.  When we apply regular multiple pulses with an interval \(\tau_{s}\) and  a pulse duration \( \Delta t \) after applying a pulse at an initial time (\(t=0\)), the quantity is written as 
\begin{eqnarray}
\langle 1 | Tr_{R} \; \rho(t) | 0 \rangle  \nonumber \\
&&\hspace{-2.5cm} =Tr_{R} \langle 1 |e^{-i L_{R}(t-(N\tau_{s}+\Delta t))} 
T_{+} [e^{-i \int_{ N \tau_{s} }^{N \tau_{s} +\Delta t} dt' L_{P,j} (t')}] \nonumber \\
&&\hspace{-2cm}  \times \{\prod_{j=0}^{N-1} e^{-i L_{R} (\tau_{s}-\Delta t)} 
T_{+} [e^{-i \int_{ j \tau_{s} }^{j \tau_{s} +\Delta t} dt' L_{P,j} (t')}]\} 
\rho(0) | 0 \rangle\;, \nonumber \\
\label{eqn:6}
\end{eqnarray}
where \(T_{+}\) is the time ordering symbol from right to left, and \( L_{\nu} \; (\nu=\{P,j\} \; or \; \{R\})\) is the Liouville operator defined as
\begin{equation}
i L_{\nu} \;\; \cdots \;\; \equiv  \ih [\ch_{\nu} \;\; , \cdots\;\;]\;,
\label{eqn:7}
\end{equation}  
and \(\ch_{P,j}\) is the Hamiltonian during the \(j\)-th pulse.

For later convenience, we introduce the hyperspace notation as \(|m\rangle \langle n|\; \to \;|mn\rangle \rangle \), \(Tr[(|k\rangle \langle l|)^\dag  \;|m\rangle \langle n|] \to \;\langle \langle kl|mn\rangle \rangle \) and  \(Tr[(|k\rangle \langle l|)^\dag  (O\;|m\rangle \langle n|)] \to \;\langle \langle kl|O|mn\rangle \rangle \) for an arbitrary hyperoperator \(O\). Using the completeness relation,
\begin{equation}
\sum_{m=0,1}\sum_{n=0,1} |  m n \rangle \rangle \langle \langle m n | =1 \;,
\label{eqn:8}
\end{equation}  
Eq.(\ref{eqn:6}) is separated into the following elements as
\begin{eqnarray}
\langle \langle k l | e^{-i \int_{t_{1}}^{t_{2}} dt' L_{P} (t')}| m n \rangle \rangle \nonumber \\
&&\hspace{-4cm} =e^{-i \omega  \{(k-l) t_{2} -(m-n) t_{1} \} }  
( \cos(\frac{\theta}{2}) \delta_{k,m} + i \sin(\frac{\theta}{2}) (1- \delta_{k,m} )) \nonumber \\
&& \hspace{-3.5cm}  \times ( \cos(\frac{\theta}{2}) \delta_{n,l} - i \sin(\frac{\theta}{2}) 
s(1- \delta_{n,l} )) 
\end{eqnarray}
with the so-called pulse area \(\theta \equiv \frac{(t_{2}-t_{1}) }{\hbar} \muv \cdot \ev\), and
\begin{equation}
\langle \langle k l | e^{-i L_{R} t} \rho_{R} | m n \rangle \rangle 
= \delta_{k,m} \delta_{l,n} \exp[-i L_{m,n} t] \rho_{R} \;.
\label{eqn:9}
\end{equation}
In Eq.(\ref{eqn:9}),  \(\rho_{R} \) is the density operator for the reservoir, and
\begin{equation}
L_{m,n}  \;\; \rho_{R} = \ch_{B,m} \;\; \rho_{R} - \rho_{R} \;\; \ch_{B,n} \;,
\label{eqn:10}
\end{equation}
with
\begin{equation}
\ch_{B,m} =\frac{2m-1}{2} ( \hbar \omega + B) +\ch_{B} \;.
\label{eqn:10a}
\end{equation}

Now we consider the case where we first apply a \(\frac{\pi}{2}\) pulse to generate a superposed spin state at an initial time (\(t=0\)) and then apply \(N\) times of \(\pi\) pulses with constant interval \(\tau_{s}\).  
Assuming the initial total system to be
\begin{equation}
\rho(0) =\rho_{R}(0) | 0 \rangle \langle 0 | 
\label{eqn:18a}
\end{equation}  
and  using Eqs. (\ref{eqn:7}) \(\sim\) (\ref{eqn:10a}), we have for even \(N\),
\begin{eqnarray}
\langle 1 | Tr_{R} \; \rho(t) | 0 \rangle \nonumber \\ 
&& \hspace{-2cm} =Tr_{R} [ \langle \langle 1 0 | e^{-i L_{R} (t-N \tau_{s}) } |  1 0 \rangle \rangle  \nonumber \\
&& \hspace{-1.5cm} \times \{\langle \langle 0 1 | e^{-i L_{R} \tau_{s} } | 0 1 \rangle \rangle 
           \langle \langle 1 0 | e^{-i L_{R} \tau_{s} } | 1 0 \rangle \rangle \}^{\frac{N}{2}}  
           \rho_{R} ] \nonumber \\
&& \hspace{-2cm} =Tr_{R} [ e^{-i L_{1,0} (t-N \tau_{s}) } 
           \{ e^{-i L_{0,1} \; \tau_{s} }  e^{-i L_{1,0} \; \tau_{s} } \}^ {\frac{N}{2}}  
           \rho_{R} ] \nonumber \\
&& \hspace{-2cm} =Tr_{R} [ e^{-i \ch_{B,1} \; (t-N \tau_{s}) } 
            \{e^{-i \ch_{B,0} \; \tau_{s}}  e^{-i \ch_{B,1} \; \tau_{s} } \}^{\frac{N}{2}}   \nonumber \\ 
  && \hspace{-1.5cm}   \times \rho_{R} \{e^{i \ch_{B,0} \; \tau_{s} } e^{i \ch_{B,1} \; \tau_{s} }\}^{\frac{N}{2}}  
            e^{i \ch_{B,0} \; (t-N \tau_{s}) }\nonumber \\   
\label{eqn:18b}
\end{eqnarray}

\begin{figure}[h]
\begin{center}
\includegraphics[scale=0.5]{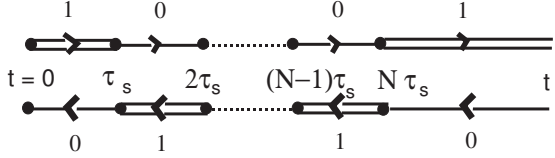}
\end{center}
\caption{The diagram for the time evolution of the reduced density operator in Eq.(\ref{eqn:18b}). }
\label{fig:fig1}
\end{figure}

In the last line of Eq.(\ref{eqn:18b}), the exponentials with exponent \(\ch_{B,1} \) and \(\ch_{B,0} \) alternately appear, which reflects the switching between states \(| 1 \rangle \) and \(| 0 \rangle \) by \(\pi\) pulses. This is schematically  drawn in Fig.\ref{fig:fig1}. The diagram consists of two parallel straight lines: the upper(lower) line is associated with  forward (backward) time evolution of \(\rho_{R}\) in Eq.(\ref{eqn:18b}). The arrow indicates the direction of time evolution. The thick (thin) line segment corresponds to the time evolution in the state \(| 1 \rangle \) (\(| 0 \rangle \)).  

We can rewrite Eq.(\ref{eqn:18b}) in the form
\begin{eqnarray}
\langle 1 | Tr_{R}  \; \rho(t) | 0 \rangle \nonumber \\
&&\hspace{-2.5cm} =(-1)^{N} \frac{i}{2} \;  e^{-i \omega \; (t-N \tau_{s}) }  
Tr_{R}[ U( \tau_s , (t-(N-1) \tau_s) )  \nonumber \\
&& \hspace{-2cm} \times  U^{\dagger} ( 0 ,\tau_s ) U^{\dagger} (  2 \tau_s , 3 \tau_s)   
\cdots U^{\dagger} ( (N-2) \tau_s, (N-1) \tau_s)   \nonumber \\
&& \hspace{-2cm} \times  \rho_{R}(0) \nonumber \\
&&\hspace{-2cm} \times  U( (N-3) \tau_s , (N-2) \tau_s) 
\cdots U( 3\tau_s , 4 \tau_s )  U( \tau_s , 2 \tau_s ) \nonumber \\
&&\hspace{-2cm} \times U^{\dagger} (0,\tau_s) ] \;,
\label{eqn:20}
\end{eqnarray}
where we use the relation 
\begin{equation}
e^{-\ih (\ch_{B}+B) (t_{1}-t_{2}))} = e^{-\ih \ch_{0} t_{1}} U(t_{1} , t_{2} ) e^{\ih \ch_{0} t_{2}}  
\end{equation}
and
\begin{eqnarray}
U(t_{1} , t_{2} ) &\equiv& T_{+} [\exp[-\ih \int_{t_{1}}^{t_{2}} dt'B(t')] ]\;, \label{eqn:17} \\
B(t) &\equiv& e^{\ih \ch_{B} t} B e^{-\ih \ch_{B} t} \;. \label{eqn:18}  
\end{eqnarray} 

Using the diagram method in Appendix A to take up to the second order of \(B(t)\) in cumulant expansion, we obtain the signal intensity for the coherently oscillating spin described by the off-diagonal element of the reduced density operator, \(\langle 1 |  Tr_{R} \; \rho(t) | 0 \rangle \),  as  
\begin{eqnarray}
I(t) &\equiv& |\langle 1 | Tr_{R} \; \rho(t) | 0 \rangle|^2 \nonumber \\
&&\hspace{-0.5cm} =\exp[-2((-1)^{N}\sum_{n=1}^{N}(-1)^{n}(4n-2) S((N-n+1)\tau_{s}) \nonumber \\
&&\hspace{0cm} + (-1)^{N-1} \sum_{n=1}^{N} 2(-1)^{n+1} S(t-n\tau_{s}) \nonumber \\
&&\hspace{0cm} +(-1)^{N} S(t))]\;,
\label{eqn:21}
\end{eqnarray}
where we introduce the second cumulant as
\begin{equation}
S(t) \equiv \int_{0}^{t} dt_{1} \int_{0}^{t_{1}} dt_{2} Re[\langle B(t_{1}) B(t_{2}) \rangle]\;. 
\label{eqn:22}
\end{equation}  
Although the formula (\ref{eqn:21}) is obtained up to the second order of \(B(t)\) in cumulant expansion, we should recall that the formula can explain the pure dephasing in many systems where spin-reservoir interaction is composed of a strong linear interaction and a relatively weak nonlinear one. This is because (1) the higher-order cumulants vanish irrespective of the coupling strength for the nondegenerate two-level system linearly interacting with the boson reservoir\cite{kotani}, and (2) they are negligible for the nonlinearly interacting system if the interaction is sufficiently weak. 

The quantity \(S(t)\) can be rewritten as
\begin{equation}
S(t)= \int_{-\infty}^{\infty} \frac{d\omega}{2\pi} J(\omega) \frac{1-\cos{\omega t}}{\omega^2} \;,
\label{eqn:23}
\end{equation}  
where \(J(\omega) \) is the power spectrum defined by 
\begin{equation}
J(\omega) \equiv \int_{-\infty}^{\infty}  dt e^{i \omega t} \langle B(t)B(0) \rangle \;.
\label{eqn:24}
\end{equation}

 The obtained general formula (\ref{eqn:21}) describes the spin dynamics under multipulse control in terms of the correlation function of the interaction between system and reservoir without specifying the form of interaction amplitude \(B\), and can be applied to various systems.

At the last part of this section, we consider the asymptotic behavior of the quantity \(S(t) \) in the long time region.
Rewriting Eq.(\ref{eqn:23}) into the form as
\begin{eqnarray}
S(t)&=& \int_{-\infty}^{\infty} \frac{d\omega}{2\pi} J(\omega) \frac{1-\cos{\omega t}}{\omega^2} \nonumber \\
&=&\int_{-\infty}^{\infty} \frac{d\omega}{2} \frac{1-\cos{\omega t}}{\pi t \omega^2} J(\omega) \; t \; , 
\label{eqn:a1}
\end{eqnarray}
we have an asymptotic form as 
\begin{equation}
S(t) \sim \frac{J(0)}{2} t \;,  
\label{eqn:a3}
\end{equation}
since
\begin{equation}
\frac{1-\cos{\omega t}}{\pi t \omega^2} \sim \delta(\omega)  \;,
\label{eqn:a2}
\end{equation}
in the long time region (\( t \gg \tau_c\)).
The relation (\ref{eqn:a3}) means that the quantity \(S(t) \) for large \(t\) is determined
by the value  \(J(0)\),  the power spectrum at \(\omega=0\).
Using the definition of \(\frac{J(0)}{2} = \frac{1}{T_{2}} \),
 we find that the general formula (\ref{eqn:21}) under the multipulse control with the long interval is reduced to
\begin{equation}
I(t) = |Tr_{R} \langle 1 | \rho(t) | 0 \rangle|^2 \sim \exp[-\frac{2}{T_{2}} t]\; .
\label{eqn:a4}
\end{equation}
The multipulse control does not work in this situation. However, as the pulse interval becomes short, the effectiveness of pulse control increases. We will show it by evaluating concrete cases in the next section.

\section{Model Calculation}
\label{sec:3} 
Now we apply the general formula obtained in the previous section to the following two cases: (1) a phenomenological two-level model where the correlation function of the interaction decays exponentially, and (2) a spin-boson model where a spin nonlinearly interacts with a boson reservoir.

\subsection{Phenomenological Two-Level Model }
In order to consider the general behavior of spin dynamics and to find model-independent results, we first apply the expression (\ref{eqn:21}) to a spin system interacting with the reservoir where correlation function exponentially decays as
\begin{equation}
\langle B (t_{1}) B(t_{2}) \rangle= \Delta^2 \exp[-|t_{1}-t_{2}|/ \tau_{c} ] \;,
\label{eqn:31}
\end{equation}
where \(\tau_{c}\) denotes the correlation time of the reservoir and \(\Delta\) is the strength of the interaction.
In this case, Eq.(\ref{eqn:22}) is written as
\begin{equation}
S(t)=-\Delta^2 \tau_c \{t-\tau_c (1-e^{-t/\tau_c}) \}  \;.
\label{eqn:32}
\end{equation}

Although the correlation function (\ref{eqn:31}) is simple, we can find various features depending on the ratio \(t/\tau_c\). In the long time region (\(t \gg \tau_c\)), the time evolution of the spin is characterized by
\begin{equation}
S(t)=-\Delta^2 \tau_c t = - \frac{t}{T_2} \;,
\label{eqn:33}
\end{equation}
where we defined the dephasing time \( T_{2} \equiv ( \Delta^2 \tau_{c} )^{-1}\). 
(Using Eq.(\ref{eqn:24}) and Eq.(\ref{eqn:31}), we have \(J(0)=2 \Delta^2 \tau_{c}= \frac{2}{T_{2}} \).) Since the memory of the detailed information on the interaction with the reservoir is lost, the time evolution becomes fully irreversible and multipulse control is not effective.  However, the multipulse control with sufficiently short pulse interval works well as shown below, because the memory effect of the reservoir plays an important role in the short time region. Equation (\ref{eqn:32}) gives the dependence of \(\ln [I(t)]\) at the pulse applied time \(t_{N} (\equiv N \tau_{s}) \) for even \(N\) in the form
\begin{eqnarray}
\ln[I(t_{N} )] \nonumber \\
&&\hspace{-1.5cm} =- \frac{2}{T_2} 
\{t_{N} + \left( 1 + e^{ - \frac{t_{N} }{ \tau_c} } - 2\,N \right) \, \tau_c \nonumber \\
&&\hspace{-1cm}  +\frac{\left( -2\,e^{\frac{\left( \tau_s  - t_{N} \right) \, } { \tau_c} } + 
       2 e^{-\frac{ t_{N} }{ \tau_c} }+ 4\,N \right) \, \tau_c}{1 + e^{\frac{ \tau_s}{ \tau_c} }} \nonumber \\
&&\hspace{-1cm} -\frac{4 \left( \,e^{\frac{ \tau_s}{ \tau_c} } + e^{-\frac{t_{N} }{ \tau_c} }  \right) \, \tau_c} 
{{\left( 1 + e^{\frac{ \tau_s}{ \tau_c} } \right) }^2} \} \;.
\label{eqn:34}
\end{eqnarray}

In the case of large \(\tau_{s}\) (\(\gg \tau_c\)), we obtain the following approximate expression
\begin{equation}
\ln[I(t_{N})] \sim  -\frac{2} {T_{2}^{e}} \; t_{N}  \;,
\label{eqn:35}
\end{equation}
introducing an effective dephasing time:
\begin{equation}
{T_{2}^{e}} \equiv \frac{T_2} {(1\, + \left( \frac{1}{N} - 2 \right) \; \frac{\tau_c}{\tau_s} )} \;.
\label{eqn:36} 
\end{equation}
Comparing Eq.(\ref{eqn:35}) and Eq.(\ref{eqn:a4}), we find that the effective dephasing time \(T_{2}^{e}\) increases with decreasing \(\frac{\tau_{s}}{\tau_{c}}\) to give suppression by the multipulse control.

\begin{figure}[h]
\begin{center}
\includegraphics[scale=0.6]{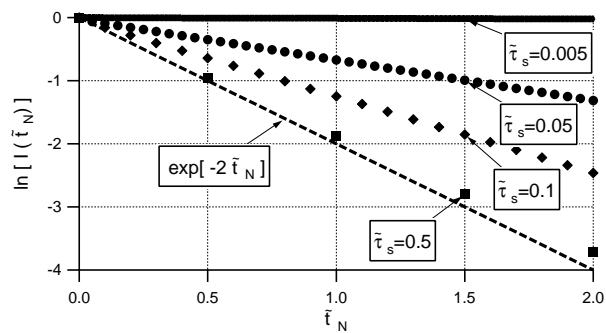}
\end{center}
\caption{The signal intensity \(\ln [ I ( {\tilde t} _{N})]\) as a function of a pulse application time for several values of the pulse interval \(\taus\). The pulse interval \(\taus\) is set to \(0.005,0.1,0.5\) for a constant correlation time (\(\tauc=0.02\)). The dashed line shows the exponential decay with \(T_{2}\) as a reference.}
\label{fig:fig2}
\end{figure}

Using the dimensionless time scale normalized by the dephasing time \(T_{2}\) as \( {\tilde \tau_{c(s)}} \equiv \tau_{c(s)}/T_{2}\) and \({\tilde t}_{N} \equiv \tau_{c(s)}/T_{2}\), we evaluate the signal intensity \(\ln [ I ( {\tilde t} _{N})]\) as a function of a pulse application time for several values of the pulse interval \(\taus\), while the correlation time \(\tauc\) is fixed to 0.02 (see Fig.\ref{fig:fig2} ).  

When  the pulse interval \(\taus \) is much larger than the correlation time (\(\taus=0.5\)), the decay is nearly determined by the pure dephasing time \(T_{2}\).  As  \(\taus \) decreases to \(0.1\), we can see that the decay of the \(\ln [ I ( {\tilde t} _{N})]\) becomes slow. This means that the pulse train recovers the phase coherence.  When the pulse interval becomes much shorter than the correlation time (\(\taus=0.005\)), we can see that the pure dephasing is highly suppressed.

\begin{figure}[h]
\begin{center}
\includegraphics[scale=0.6]{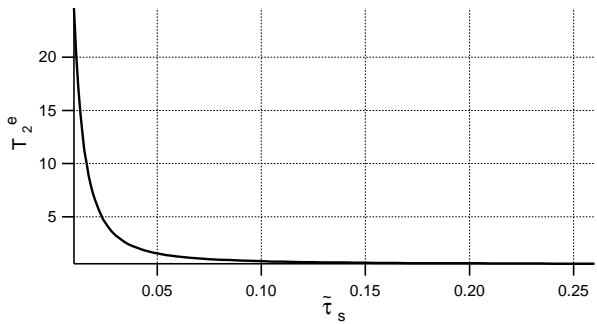}
\end{center}
\caption{The pulse-interval dependence of the effective dephasing time \(T_{2}^{e}\) .}
\label{fig:fig3}
\end{figure}

Evaluating numerically  effective dephasing time \(T_{2}^{e}\) from the decay time at sufficiently large time \({\tilde t}=10\), we show  the pulse-interval dependence of the effective dephasing time \(T_{2}^{e}\) in Fig.\ref{fig:fig3}.  We find \(T_{2}^{e}\) becomes large for  sufficiently small pulse interval and monotonously decreases to an asymptotic value \(0.5\).

\begin{figure}[h]
\begin{center}
\includegraphics[scale=0.6]{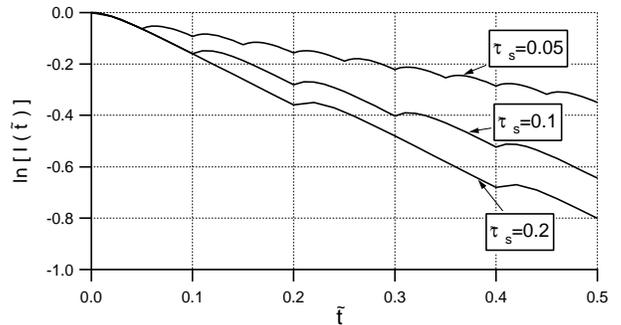}
\end{center}
\caption{Time evolution of the signal intensity \(\ln [ I ( {\tilde t})]\) under the multipulse control. The pulse interval \(\taus\) is set to \(0.05,0.1,0.2\) for a constant correlation time (\(\tauc=0.02\)). }
\label{fig:fig4}
\end{figure}

Let us consider in more detail the physical origin why the dephasing is suppressed by the pulse-train control. The general formula (\ref{eqn:21}) gives the time evolution of the system between the pulses as well as the values on the pulse-applied time.  Figure \ref{fig:fig4} shows the time dependence of \(\ln [ I ( {\tilde t})]\) during the multipulse control for \(\taus=0.05, 0.1\) and \(0.2\) with a constant correlation time (\(\tauc=0.02\)).   We can see the decay recovers between the pulses. This is because the applied \(\pi\) pulses cause the time reversal for spin dynamics.  The degree of recovery reflects to what extent the memory of the interaction with the reservoir remains, which depends on the ratio \(\frac{\tau_{s}}{\tau_{c}}\).

The memory effects of the reservoir in optically excited systems have 
been extensively studied during the past couple of decades as the 
non-Markovian optical problem\cite{aihara1,aihara3,agarwal3,fainberg,vogel,tchenio,nibbering,bigot,lavoine,saikan,hamm}.  
We should stress that the dephasing control by a \(\pi\)-pulse train discussed 
here is one of the most typical and clear phenomena arising from the 
finiteness of the correlation time.

\subsection{Nonlinear Spin-Boson Model}
Applying the formula (\ref{eqn:21}) to the case where the correlation of the interaction between system and reservoir is written in an exponential form, we have quantitatively evaluated \(\ln [ I ( {\tilde t})]\)  and discussed the physical origin of the suppression of the pure dephasing.
Next we consider a microscopic case where the pure dephasing occurs by the interaction between a \(\frac{1}{2}\) spin and a reservoir which consists of bosons.  While the standard spin-boson model\cite{Caldeira} has been often used to describe pure dephasing by linear spin-reservoir interaction, there exist many dephasing phenomena which cannot be explained without non-linear spin-reservoir interaction.  A typical example of this situation is the pure dephasing phenomena of a localized electron in a solid.  The dephasing phenomena, which are originated from the interaction with phonons, cannot be described by assuming the interaction to be linear where absorption and emission of one phonon occurs.  As shown in sec.\ref{sec:2}, the long time behavior is described by the value of the spectrum \(J(\omega)\) at \(\omega=0\) (see Eq.(\ref{eqn:a3})).  In the case of linear interaction, the value \(J(0)\) corresponds to the interaction with the acoustic phonon at the wave number \(k=0\), which corresponds to the translation of the whole solid without distortion.  Since such a motion does not affect the electron, the pure dephasing cannot be described with the standard spin-boson model\cite{Caldeira}. We need to consider a nonlinear interaction with the reservoir.  The reason why the linear spin-boson model cannot generally describe the pure dephasing is given in Appendix B.

In this paper, assuming the reservoir to be composed of the infinite number of bosons
\begin{equation}
\ch_{B}=\sum_{l} \hbar \omega_{l} b_{l}^{\dagger} b_{l} \;,
\label{eqn:42a}
\end{equation}
in an equilibrium state with temperature \(T\),
we take into account the quadratic interaction between the spin and reservoir as
\begin{eqnarray}
\ch_{SB} =\hbar S_{z}B \nonumber \\
&& \hspace{-1.5cm} =\frac{\hbar}{2} S_{z} \sum_{l} \sum_{m} h_{lm} \sqrt{ \omega_{l} \omega_{m}} (b_{l}+b_{l}^{\dagger})(b_{m}+b_{m}^{\dagger})\;.\nonumber \\
\label{eqn:41}
\end{eqnarray} 
The interaction Hamiltonian (\ref{eqn:41})  describes simultaneous absorption and/or emission of two phonons with frequencies \(\omega_{l}\) and \(\omega_{m}\) where interaction strength is \( h_{lm} \).  The term \(b_{l}^{\dagger}b_{m}\) in the above interaction Hamiltonian causes the emission of a phonon with frequency \(\omega_{l}\) and the absorption of a phonon with frequency \(\omega_{m}\) at the same time.  The energy change of this process is \(\hbar( \omega_{l}-\omega_{m})\), and therefore it contributes 
to \(J(\omega=0)\) as shown below. This is the reason why the nonlinear interaction is essential when we consider the pure dephasing phenomena.

The quadratic interaction gives the correlation function in Eq.(\ref{eqn:24}) of the form
\begin{eqnarray}
\langle B(t) B(0) \rangle \nonumber \\
&& \hspace{-2.5cm} =\sum_{l} \sum_{m} h_{lm}^2 \omega_{l} \omega_{m}
\{ (n(\omega_{l})+1) (n(\omega_{m})+1) \; e^{-i (\omega_{l}+\omega_{m}) t} \nonumber \\
&& \hspace{-1.5cm} +2 n(\omega_{l}) (n(\omega_{m})+1) \; e^{i (\omega_{l}-\omega_{m}) t} \nonumber \\
&& \hspace{-1.5cm} + n(\omega_{l}) n(\omega_{m}) \; e^{i (\omega_{l}+\omega_{m}) t} \} \;,
\label{eqn:42}
\end{eqnarray} 
where \(n(\omega)\) is the boson distribution function:
\begin{equation}
n(\omega)=\frac{1}{e^{\hbar \omega/k_{B} T}-1} \;.
\label{eqn:43}
\end{equation}

We rewrite the summation over the label of the boson  in Eq.(\ref{eqn:42}) into an integral form to give the power spectrum \(J(\omega)\) in Eq.(\ref{eqn:24}) by using the following relation, 
\begin{eqnarray}
\sum_{l} \sum_{m} h_{lm}^2  f(\omega_{l},\omega_{m})  \nonumber \\
&&\hspace{-3cm} =\sum_{l} \sum_{m} h_{lm}^2  \delta(e-\omega_l) \delta(e'-\omega_m) f(e,e')  \nonumber \\
&&\hspace{-3cm}  =\int_{0}^{\infty}  de \int_{0}^{\infty}  de' h(e) h(e') f(e,e')  
\label{eqn:44}
\end{eqnarray} 
where \(h(\omega)\) is a spectral density of the coupling function for the spin-reservoir interaction and \(f (\omega,\omega') \) is an arbitrary function of \(\omega\) and  \(\omega'\). 

When the spectral density of the coupling function is assumed to be  a Gaussian distribution with mean frequency \(\omega_{p}\) and variance \(\gamma_{p}\);
\begin{equation}
h(e) \equiv \frac{s }{\sqrt{\pi} \gamma_{p} } \exp(-\frac{(e-\omega_{p})^2}{\gamma_{p}^2}) \;, 
\label{eqn:45}
\end{equation}
we have
\begin{eqnarray}
J(\omega)     \nonumber \\
&& \hspace{-1cm} =2 \pi \int de h(e) \; \{ e( \omega-e ) (n(e)+1) (n(\omega-e )+1) h(\omega-e ) \nonumber \\
&& \hspace{-0.5cm} - 2 e ( \omega-e ) n( -( \omega-e ) ) (n(e)+1) h(-(\omega-e)) \nonumber \\ 
&& \hspace{-0.5cm} -  e ( \omega+e ) n(e) n(-( \omega+e ))  h(-(\omega+e)) \} \;.
\label{eqn:46}
\end{eqnarray} 

\begin{figure}[h]
\begin{center}
\includegraphics[scale=0.6]{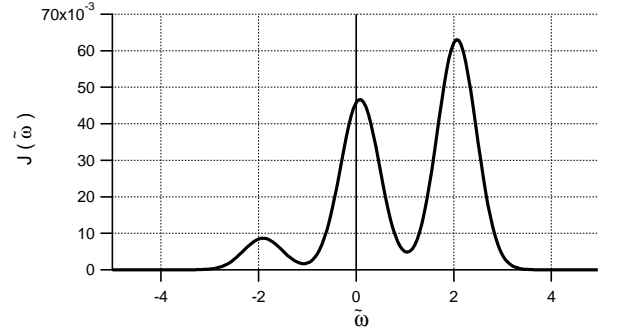}
\end{center}
\caption{Power spectrum \(J( {\tilde \omega} )\) for \( {\tilde \gamma_p}=0.4\), \(S_{Q}=1/40\), and \(\hbar \omega_{p}/k_{B} T =1\). } 
\label{fig:fig5}
\end{figure}

Figure \ref{fig:fig5} shows \(J( {\tilde \omega} )\) for \( {\tilde \gamma_p} =0.4\), \(S_{Q}\equiv s^2 =1/40\), and \(\hbar \omega_{p}/k_{B} T =1\) where we have defined as \({\tilde \omega} \equiv \omega/\omega_{p}\) and \({\tilde \gamma_p} \equiv \gamma_p/\omega_p\).  
We can see three peaks: The right (left) peak at \( {\tilde \omega}=2 (=-2)\) in Fig.\ref{fig:fig5} corresponds to the emission (absorption) of two bosons by the spin. The center peak  at  \({\tilde \omega}=0\) represents the simultaneous boson absorption and emission, which causes the pure dephasing. This is because the quantity \(S(t)\) is described as \(S(t) \sim \frac{J(0)}{2} t \) in the long tine region, and the quantity \(\frac{J(0)}{2}\) gives the dephasing rate for the dynamics(see  Eq.(\ref{eqn:a3}) and Eq.(\ref{eqn:a4})). 

\begin{figure}[h]
\begin{center}
\includegraphics[scale=0.6]{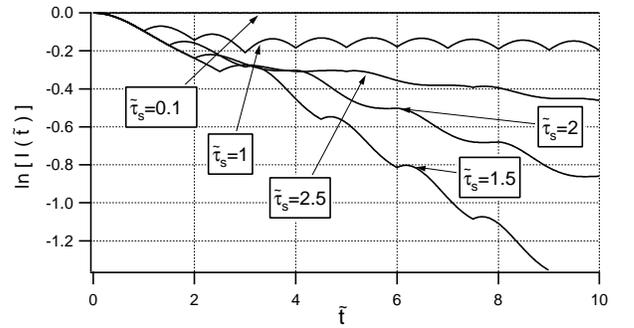}
\end{center}
\caption{Logarithmic display of intensity \(\ln [ I ( {\tilde t} )]\) for several values of pulse interval \({\tilde \tau_{s} } =0.1,1.5,2,2.5\) in the case of  \( {\tilde \gamma_p}  =0.4\), \(S_{Q}=1/40\), and \(\hbar \omega_{p}/k_{B} T =1\). }
\label{fig:fig6}
\end{figure}

Now we discuss the effectiveness of the multipulse control in suppression of the pure dephasing.  In Fig.\ref{fig:fig6}, we show \(\ln [ I ( {\tilde t} )]\) for the same conditions as in the Fig.\ref{fig:fig5}, using a definition as \({\tilde t} \equiv \omega_{p} t\). The decay is well suppressed for small pulse interval \( {\tilde \tau_{s} } (\equiv \omega_{p} \tau_{s})=0.1\), which is consistent to the results in the previous subsection.  However, a surprising result is obtained in increasing the pulse interval: As the interval increases to \({\tilde \tau_{s} } =1.5\), the degree of the suppression becomes worse, but a significant recovery is seen for \( {\tilde \tau_{s} } =2\) and \(2.5\).

\begin{figure}[h]
\begin{center}
\includegraphics[scale=0.6]{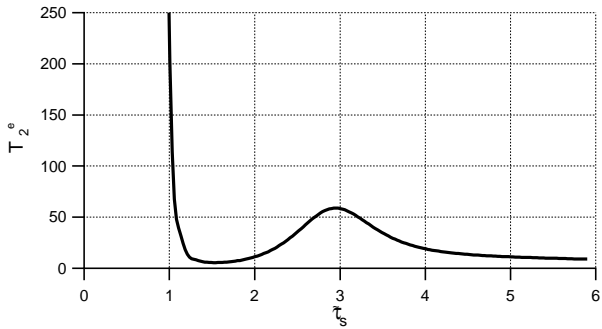}
\end{center}
\caption{The pulse interval dependence of the effective dephasing time \(T_{2}^{e}\) for the non-linearly interacting spin-boson model.}
\label{fig:fig7}
\end{figure}

Next we evaluate the pulse-interval dependence of the effective dephasing time \(T_{2}^{e}\) which is obtained by the decay time at \({\tilde t}=45\).  Figure \ref{fig:fig7} shows that the characteristic feature inherent to the dynamical behavior of boson system.  When the pulse interval is near to \(\frac{\pi}{\omega_{p}}\), the more effective suppression of  pure dephasing is caused by resonance enhancement between the oscillatory behavior of boson system and the pulse train.  This should be contrasted with the phenomenological case where \(T_{2}^{e}\) monotonously decreases with increasing the pulse interval \(\tau_{s}\) (see Fig.\ref{fig:fig3}).  

\section{Concluding Remarks}
\label{sec:4} 
We have obtained a general formula (\ref{eqn:21}) which describes the time evolution of spin system under pulse-train control. The formula is expressed with the correlation function for arbitrary interaction amplitude \(B\) which causes dephasing.  This enables us to clarify how the suppression of pure dephasing depends on the pulse interval and the characteristic time of the interaction.   We have applied the general formula to two examples: (1) a phenomenological two-level model where the correlation function of the interaction decays exponentially, and (2) a spin-boson model where a spin nonlinearly interacts with a boson reservoir.  In the case (1), the analytic form for the intensity is evaluated from the off-diagonal element of the reduced density operator at pulse applied times. We define effective pure dephasing time \(T_{2}^{e}\) from this form and we find that \(T_{2}^{e}\) effectively increases by decreasing the ratio of the pulse interval to the correlation time.  Moreover, the analytic expression of the time evolution of the intensity between pulses enables us to generally discuss the mechanism of the suppression of dephasing by multipulse control.  
In the case (2), we also found the suppression of the pure dephasing.  However, special attention should be paid to the fact that the extent of the suppression does not monotonously decrease as the pulse interval increases, which is contrasted with the phenomenological case (1). When the pulse interval is near to the half of the average period of boson system, pure dephasing is effectively suppressed by resonance  between the oscillatory behavior of boson system and the pulse train.  This coherence recovery shows us that the microscopic mechanism can be used for effective suppression. 

\begin{acknowledgements}
This study is supported by the Grant in Aid for Scientific Research from the Ministry of Education, Science, Sports and Culture of Japan. The authors are deeply grateful to Dr. L. Viola for her thoughtful comments and suggestions.

\end{acknowledgements}

\appendix
\section{Diagram Method}
\begin{figure}[h]
\begin{center}
\includegraphics[scale=0.3]{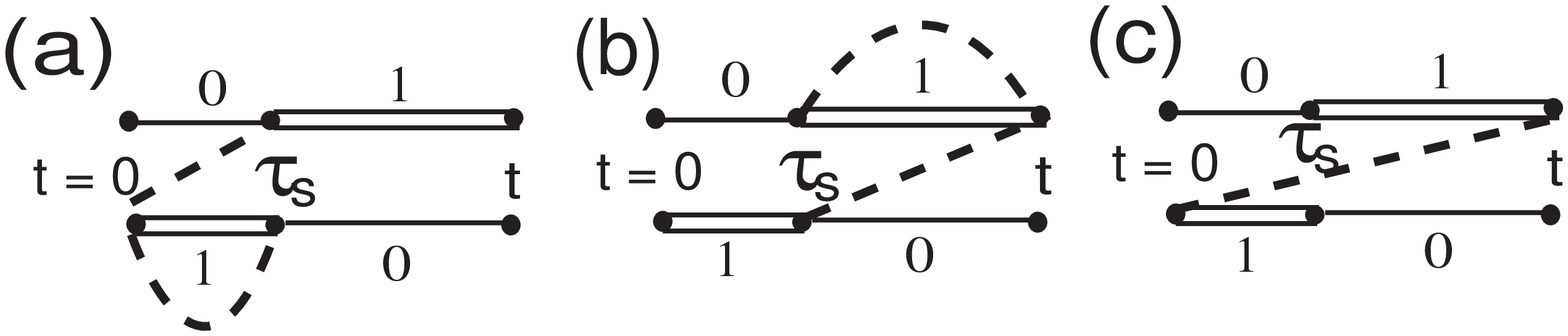}
\end{center}
\caption{Diagrams for Eq.(\ref{eqn:ap1}). Diagram (a) corresponds to \(2 S(\tau_s) \), (b) corresponds to \(2 S(t-\tau_s) \), and (c) corresponds to \( -S(t) \). }
\label{fig:fig8}
\end{figure}

In order to obtain Eq.(\ref{eqn:21}), we expand the time-ordered exponential in Eq.(\ref{eqn:20}) and take up to the second order of \(B(t)\).  In the case of two-pulse excitation 
, for the most simple case where \( \pi \) pulse is applied once at \( t=\tau_s \) after  a \( \frac{\pi}{2} \) pulse at \( t=0 \), the intensity \( I(t) \) is obtained as\cite{aihara1,aihara2}
\begin{equation}
I(t) = \exp [-2 ( 2 S(\tau_s) + 2 S(t-\tau_s) -S(t) )] \; .
\label{eqn:ap1}
\end{equation}
Using the fact that the exponent in Eq.(\ref{eqn:ap1}) is composed of the function \(S\) with three kinds of time difference, \( \tau_s,  t-\tau_s \), and \( t \), we can write a diagram as in Fig.\ref{fig:fig8}.     

In this diagram, the dots are placed at the pulse applied times 
and the observation time t. The broken lines are connected between the time points that are the ends of the broad lines. The coefficient of the function \(S\) can be determined by the number of the broken line \( N_{b} \) and the number of the time points \( N_{t} \) included between the broken line as \((-1)^{ N_{t} } N_{b}\). 

\begin{figure}[h]
\begin{center}
\includegraphics[scale=0.25]{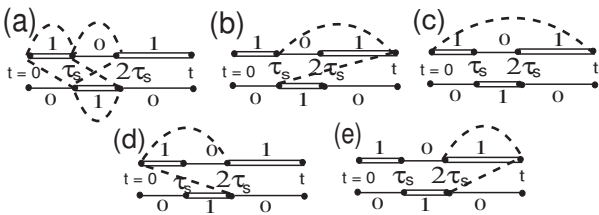}
\end{center}
\caption{Diagrams for Eq.(\ref{eqn:ap2}). Diagram (a) corresponds to \(6 S(\tau_s) \), (b) \(-2 S(t-\tau_s) \), and (c) \( S(t) \), (d) \(-2 S(2 \tau_s) \) , and (e) \(2 S(t-2 \tau_s)\). }
\label{fig:fig9}
\end{figure}

For application of \( \pi \) pulses in two times, we obtain the intensity as
\begin{eqnarray}
I(t) &=& \exp [-2 ( 6 S(\tau_s) - 2 S(t-\tau_s) +S(t) \nonumber \\
&& \hspace{1cm} - 2 S(2 \tau_s) + 2 S(t-2 \tau_s)  )] \; .
\label{eqn:ap2}
\end{eqnarray}
The diagram is written in Fig.\ref{fig:fig9}.     

These diagram rules enable us to obtain a general formula 
Eq.(\ref{eqn:21}) for an arbitrary number of pulses.  It is also  noted 
that the diagram method is a powerful tool to evaluate more general cases 
with irregular pulse interval and pulse area. 

\section{Note on the conventional linear spin-boson model}
The pure dephasing has been explained with the spin-boson model\cite{Caldeira} where a \(\frac{1}{2}\) spin linearly interacts with a reservoir which consists of bosons as,
\begin{equation}
\ch_{SB} = \hbar S_{z} B = \hbar S_{z} \sum_{k} (g_{k}^{*} b_{k}+ g_{k} b_{k}^{\dagger})\; . 
\label{eqn:b1}
\end{equation} 
In Eq.(\ref{eqn:b1}),  \(g_{k}\) is the coupling strength between the spin and the \(k\)-th boson of the reservoir.
For this model, the correlation function in Eq.(\ref{eqn:24}) is obtained as
\begin{equation}
\langle B(t) B(0) \rangle =\sum_{k} |g_{k}|^{2} 
\{ (n(\omega_{k})+1) e^{-i \omega_{k} t} + n(\omega_{k}) e^{i \omega_{k} t} \} \;, \label{eqn:b2}
\end{equation} 
which gives the power spectrum of the form,
\begin{eqnarray}
J(\omega) &=& 2 \pi \{ I(\omega) (n(\omega)+1) \theta(\omega) + I(-\omega) n(-\omega) \theta(-\omega)\} \;.  \nonumber \\
\label{eqn:b3}
\end{eqnarray} 
Here we have introduced the spectral density of the coupling function \(I(\omega)\) as
\begin{equation}
\sum_{k} |g_{k}|^{2} \delta(\omega-\omega_{k}) \cdots \equiv  \int_{0}^{\infty} d \omega I(\omega) \cdots \;.
\label{eqn:b4}
\end{equation} 
Using Eq.(\ref{eqn:23}), we obtain the second cumulant \(S(t)\) in the form
\begin{equation}
S(t) = \int_{0}^{\infty} d \omega I(\omega) (2 n(\omega)+1) \frac{1-\cos{\omega t}}{\omega^2} \;.
\label{eqn:b5}
\end{equation} 
This corresponds to \(\Gamma_{0}(t)\) of Eq.(17) in\cite{llyod1}, which describes the decay of the off-diagonal element of the density operator by the interaction with the reservoir.  In the long time region, the quantity \(S(t)\) goes to 
\begin{equation}
S(t) \sim \frac{J(0)}{2} t =\frac{\pi [I(0)(2 n(0)+1)]}{2} t \equiv \frac{t}{T_2} \;,  
\label{eqn:b6}
\end{equation}
which shows us that the value \(J(0)\) deteremines the long time behavior.

Now we consider the spectral density \(I(\omega)\) of the reservoir to be the one defined by Caldeira and Leggett\cite{Caldeira} as in previous studies\cite{llyod1,ban}, 
\begin{equation}
I(\omega)=\alpha \omega^n e^{-\omega/\omega_{c}} \;,
\label{eqn:b7}
\end{equation} 
where \(\alpha\) determines the coupling strength with the reservoir and an integer \(n\) depends on details of the system- reservoir interaction\cite{Caldeira}. 
Evaluating \(T_2\) with Eqs.(\ref{eqn:b6}) and (\ref{eqn:b7}), we have
\begin{eqnarray}
T_2  = \left\{ {\begin{array}{*{20}c}
   {\frac{2 \hbar} {\pi k_{B} T} \,\,\,(n = 1)}  \\
   { \,\,\infty \,\,\,\,\,\,\,\,\,(n \ne 1)}  \\
\end{array}}
 \right.\\\nonumber
\label{eqn:b8}
\end{eqnarray}
which shows that the pure dephasing is absent except for the ohmic dissipation case \((n=1)\).  It is remarked that the model can describe the pure dephasing only for the ohmic dissipation case.

%
%

\end{document}